\RequirePackage{fix-cm}
\documentclass[smallextended]{svjour3}
\smartqed
\usepackage{graphicx}
\usepackage{subfigure}
\usepackage{amssymb,amsmath,amsfonts}
\usepackage{color}
\usepackage{epstopdf}
\usepackage{mathptmx}
\usepackage{diagbox}
\usepackage{graphics}
\usepackage{enumerate}
\usepackage{amsxtra}
\usepackage{subfigure}
\usepackage{hyperref}
\usepackage{algorithm}
\usepackage{algorithmic}

\begin{document}

\title{Improving Vertical Positioning Accuracy with the
Weighted Multinomial Logistic Regression Classifier}

\author{Yiyan Yao         \and
        Xin-long Luo \textsuperscript{$\ast$}
}

\institute{Yiyan Yao \at
              School of Information and Communication Engineering, Beijing University
              of Posts and Telecommunications, P. O. Box 101, Xitucheng Road  No. 10,
              Haidian District, Beijing China 100876 \\
              \email{yaoyiyan@bupt.edu.cn}           
           \and
           Xin-long Luo \at
              Corresponding author. School of Information and Communication Engineering, Beijing University
              of Posts and Telecommunications, P. O. Box 101, Xitucheng Road  No. 10,
              Haidian District, Beijing China 100876 \\
              Tel.: +086-13641229062\\
              \email{luoxinlong@bupt.edu.cn}           
}

\date{Received: date / Accepted: date}

\maketitle

\begin{abstract}
  In this paper, a method of improving vertical positioning accuracy with the
  Global Positioning System (GPS) information and barometric pressure values is
  proposed. Firstly, we clear null values for the raw data collected in various
  environments, and use the 3$\sigma$-rule to identify outliers. Secondly, the
  Weighted Multinomial Logistic Regression (WMLR) classifier is trained to obtain
  the predicted altitude of outliers. Finally, in order to verify its effect,
  we compare the Multinomial Logistic Regression (MLR) method, the WMLR method,
  and the Support Vector Machine (SVM) method for the cleaned dataset which is
  regarded as the test baseline. The numerical results show that the vertical
  positioning accuracy is improved  from  5.9 meters (the MLR method), 5.4 meters
  (the SVM method) to 5 meters (the WMLR method) for 67\% test points.
\end{abstract}


\section{Introduction}
\label{INTRO}
In recent years, the performance of the Global Positioning System (GPS) is
excellent in outdoor environments \cite{RGG2015}. When users are outdoors,
their locations can be obtained accurately through GPS. However, the GPS
signals are blocked by the buildings and other obstacles, which result in
large indoor positioning errors. Thus, the indoor positioning accuracy is
often challenged, especially in the vertical direction. In the meantime, the
space that we are living in is filled with many high-rise buildings and our most
activities are indoors. Considering the practical requirement and the poor
indoor positioning performance, researchers have tried many methods to improve
the vertical positioning accuracy, such as the WiFi-based localization technology
\cite{DZYXKY2018,LLH2017,ZHLJX2016} and the barometer-based positioning
technology \cite{XWJC2015}.

\vskip 2mm

On the other hand, the GPS chip has been embedded in the most mobile terminals,
which provides the location and timing information such as time, latitude,
longitude, speed and altitude. Therefore, based on the GPS information, many
researchers put forward some effective methods to improve the positioning
accuracy of the low-cost GPS about 4 meters to 10 meters in several experiments
\cite{IK2014}. Huang and Tsai propose an approach to calibrate
the GPS position by using the context awareness technique from the pervasive
computing and improve the positioning accuracy of GPS effectively \cite{HT2008}.
The machine learning techniques are applied to assess and improve the GPS
positioning accuracy under the forest canopy in \cite{ORMMS2011}.

\vskip 2mm

In this paper, we provide another machine learning technique
\cite{ALTMY2018,AASMGK2019,AZISYDML2019} based on the Multinomial Logistic
Regression (MLR) method \cite{KS2016,MGB2008} for the vertical positioning
problem. The research data are measured by many different user equipments and
provided by Huawei Technologies Company, some data of which include the GPS
three-dimensional information and the barometric pressure values, and
Some data of which miss the GPS information or the barometric pressure values.
We preprocess the research data firstly. Consequently, we identify the abnormal
data with the $3\sigma$-rule and clear them. Meanwhile, some noises arise from
the inaccurate data records and the different reference standards of different
kinds of user equipments. These intrinsic noises lead to the poor distribution
law between the air pressure and the corresponding altitude. In order to overcome
these noise effects, we convert this vertical positioning problem into a
classification problem and revise the weighted MLR method to improve its vertical
positioning accuracy. Finally, in order to verify the effect of the Weighted
Multinomial Logistic Regression (WMLR) method, we compare the MLR method, the
WMLR method, and the Support Vector Machine (SVM) method \cite{CL2001,CL2013,CTS2017}
for this vertical positioning problem. The numerical results show that the vertical
positioning accuracy of the cleaned data is improved from  5.9 meters (the MLR method),
5.4 meters (the SVM method) to 5 meters (the WMLR method) for 67\% test points.

\vskip 2mm

The rest of the paper is organized as follows. In section \ref{RELATED}, some
related works are discussed. In section \ref{SEC:1}, we describes the methodology
of the data cleaning, the outlier detection and the data correction based on the
WMLR classifier. In section \ref{RESULT}, we describe the data source and compare
the MLR method, the WMLR method and the SVM method for the cleaned data which is
regarded as the test baseline. The promising numerical results are also reported.
Finally, some conclusions and the further works are discussed in section \ref{CONCLUSION}.

\vskip 2mm

\section{Related works} \label{RELATED}

In the field of improving the indoor vertical positioning accuracy, many
studies have been conducted. The related works can be roughly divided into two
categories: the Received Signal Strength Strength Indication (RSSI) based methods
and the barometric pressure based methods.

\vskip 2mm

The RSSI of the Wi-Fi and the cellular network based methods use the
collected the RSSI and build the database of the fingerprints for the floor
positioning \cite{BSHL2016,WLSL2014,ZLC2012}. Some researchers consider the
locations of the Wi-Fi access points to determine the floor \cite{GBRB2014}.
In \cite{BSHL2016}, the experimental data are collected from one or two buildings
and the collecting device is fixed. They use the collected RSSI information and
the pressure data to estimate the floor. In those papers, since the RSSI
information is local, when the experimental environment changes, the training 
data need to be collected by hand and the discriminant parameters need to be 
trained again.

\vskip 2mm

Since there are many Wi-Fi access points distributed in a crowded
indoor environment and the wall cannot completely obstruct the signals, the signal
interference and fluctuation of different floors will result in the inaccurate
estimation. Some researchers propose the barometric altimetry for the floor
determination. In \cite{XWJC2015}, Xia et al. give a method based on the
multiple reference barometers for the floor positioning in buildings and their
method can give an accurate floor level. The disadvantages of their method
are that the height thresholds should be given in the floor determination and
they are sensitive to the local pressure conditions.

\vskip 2mm

In \cite{CTS2017}, Chriki et al. use the SVM method based on the RSSI
measurements for the zoning localization problem. In \cite{ALTMY2018}, Adege et al.
propose an outdoor and indoor positioning method based on the hybrid of SVM
and deep neural network algorithms according to the RSSI of the Wi-Fi.
Since the SVM method only considers the support vector and the few points
which are most relevant are used to make the classification, its classification
result may be ineffective when the level of noise is high. The positioning method
based on the deep neural network \cite{ALTMY2018,HLL2017} requires a very large
amount of data to perform better than other techniques, and it requires expensive
GPUs and multiple devices to train complex models. The MLR method considers all
training data points which smooth the noise such that the MLR method can handle
the high level of noise of the training data. Furthermore, the MLR method can be
used to handle the large scale problem \cite{K2019}. Therefore, in consideration
of the performance gain of the weighted positioning algorithm \cite{LLH2017}, we
choose the MLR method with the weighted technique as the vertical positioning method
based on the GPS and barometric pressure information of the user equipments.

\section{The methodology} \label{SEC:1}

\vskip 2mm

Our positioning method is composed of several stages, including the data
cleaning, the outlier detection, the data correction and the prediction of
vertical altitude for the test feature vector. We described these procedures
in the following subsections.

\subsection{Data cleaning}  \label{SECTDATACLEAN}

\vskip 2mm

\begin{figure}[htbp]
\centering
\includegraphics[width=9cm]{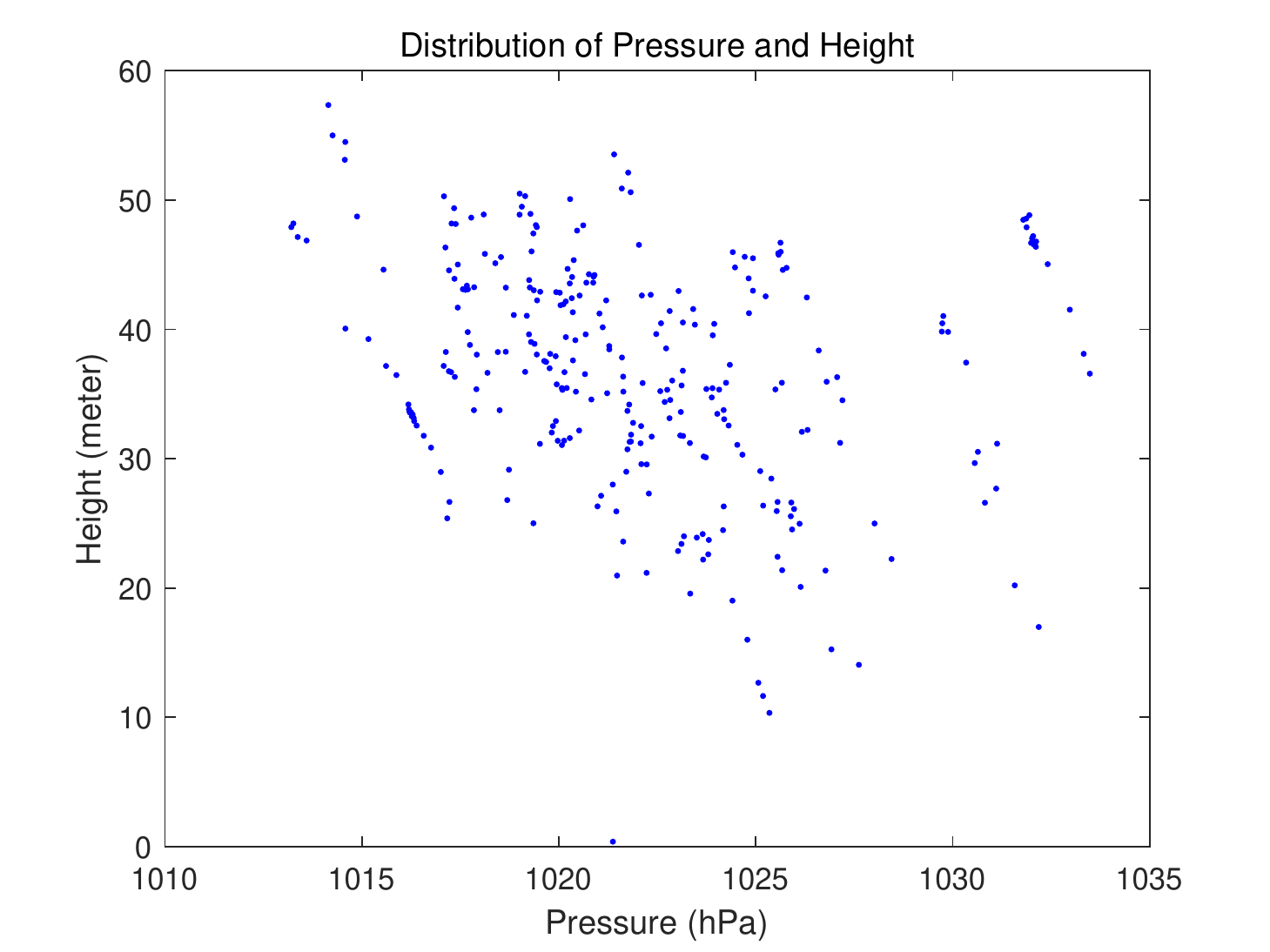}
\caption{The distribution of the pressure values and the corresponding altitudes.}
\label{Figure2}
\end{figure}

The raw dataset is measured at different places with different user equipments.
In the dataset, many data miss the air pressure values due to some mobile devices
without the barometers. We delete these data of the missing air pressure values
firstly. Additionally, there are some abnormal data which deviate too far from
the average value of the dataset and it is shown as follows. Assume that an average
sea level pressure is 1013.25 hPa and the corresponding temperature is
15$^\circ$C, then the air pressure value and its corresponding altitude have
the following relationship \cite{ZF2014}:
\begin{align}
  h=44330.8-4946.54p^{0.1902632},   \label{BAROHEIFORMULA}
\end{align}
where the unit of altitude $h$ is meter, and the unit of the air pressure value
$p$ is hpa. From formula \eqref{BAROHEIFORMULA}, it is not difficult to find
that the barometric pressure value and the corresponding altitude
are the inverse relationship. However, from Fig. \ref{Figure2}, we find that the
distribution between the air pressure values and the corresponding altitudes of
the given data is irregular. Therefore, we conclude that there exists the data drift
in the given real test data. Thus, we use the 3$\sigma$-rule to exclude the
abnormal data as follows \cite{W2004}:
\begin{align}
   X \; \text{is thrown away when} \; |X - \mu| \ge 3 \sigma, \nonumber
\end{align}
where the mean $\mu$ and the standard deviation $\sigma$ are computed by the 
following formula:
\begin{align}
 \mu =\frac{1}{n} \sum_{i = 1}^{n} X_{i} \; \text{and} \;
 \sigma = \sqrt{\frac{1}{n-1}\sum_{i = 1}^{n}(X_{i}-\mu)^2}. \nonumber
\end{align}
After performing the 3$\sigma$-rule, we eliminate the large deviation data
and the $99.73\%$ data are retained.

\vskip 2mm

\vskip 2mm

\subsection{Outlier detection} \label{SECTOUT}

\vskip 2mm

\begin{figure}[htbp]
  \centering
  \includegraphics[width=7cm]{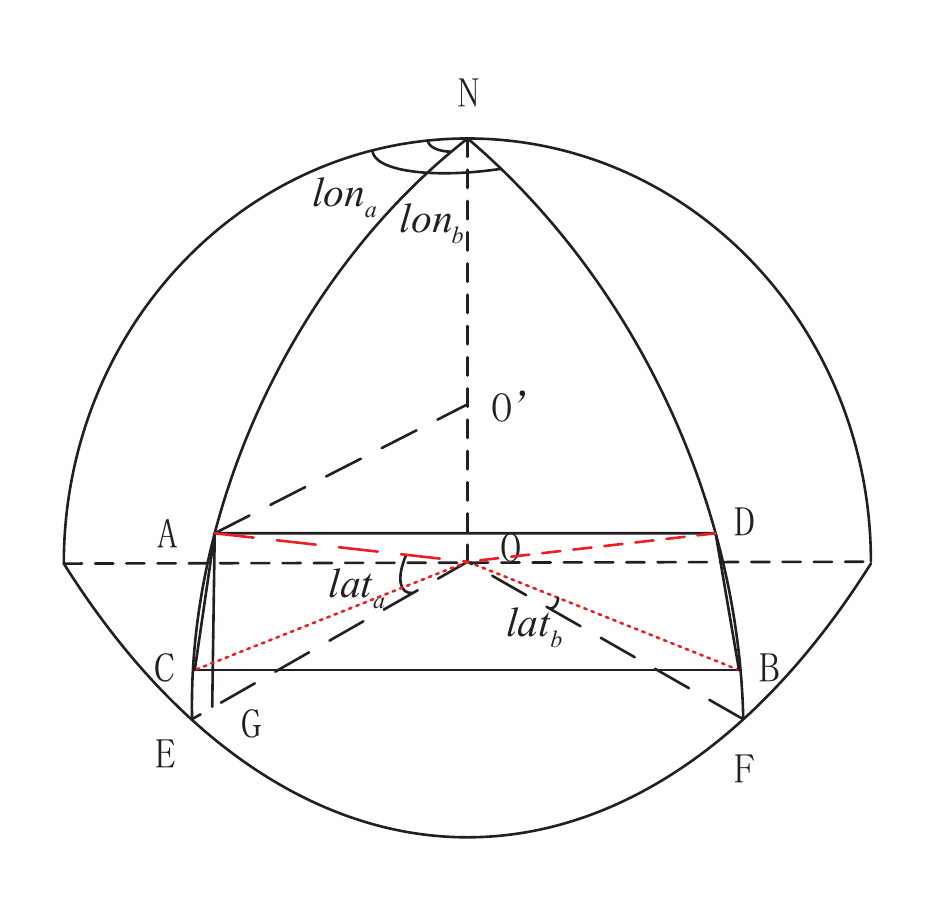}
  \caption{The diagram of two points in a three-dimensional space.}
  \label{Figure 3}
\end{figure}

In subsection \ref{SECTDATACLEAN}, we have cleaned away the abnormal data which
deviate too much from the dataset. However, there are still some outliers.
An outlier is a point which differs significantly from the other points in a subdataset
measured by the same device in a short time. We use the spherical distance computed
by the haversine formula \cite{S1984} to identify the outlier. The haversine formula
is illustrated by Fig. \ref{Figure 3} and calculates the spherical distance between
the two points $A(lon_a,\, lat_a)$ and $B(lon_b, \, lat_b)$ with the coordinate
$(longitude, \, latitude)$ as follows:
\begin{align}
  d_{AD} & = 2Rsin(\Delta lon/2)cos(lat_a), \nonumber \\
  d_{CB} & = 2Rsin(\Delta lon/2)cos(lat_b), \label{HAVFOM} \\
  d_{AB} & = 2R\left|sin^2 \left(\frac{\Delta lat}{2}\right)
    + cos(lat_a) cos(lat_b) sin^2 \left(\frac{\Delta lon}{2}\right)\right|^{\frac{1}{2}},
    \nonumber
\end{align}
where $\Delta lon=lon_{b}-lon_{a}$, $\Delta lat=lat_{b}-lat_{a}$, and $R$ is
the radius of the Earth.

\vskip 2mm

\vskip 2mm

Consequently, we estimate the diameter of a subdataset as follows:
\begin{align}
   d_{max} = \bar{v} \times t,\label{MAXDIST}
\end{align}
where $\bar{v}$ is the mean velocity, and $t$ is the total measuring time of
the subdataset. On the other hand, each point has a distance vector with other
points. If over $50\%$ elements of the distance vector are greater than
$d_{max}$, we regard this point as an outlier.

\vskip 2mm

\subsection{Data correction} \label{DATACORREC}

\vskip 2mm

In this subsection, we describe the procedure of data correction and it is also
the key step of our positioning method. This step is to predict the relatively
accurate altitudes of the outliers. As mentioned in section \ref{SECTOUT}, the
data distribution is roughly similar when the data are measured by the same
device. Under this assumption, the altitudes of the subdataset are
classified into different classes (labels). Thus, we encounter the multi-class
classification problem.

\vskip 2mm

\subsubsection{The multi-class classification problem} \label{SECPROBCONVER}

\vskip 2mm

The outliers of the subdataset have been found with the method in section
\ref{SECTOUT}. Thus, we select the data except outliers as a training dataset.
The input training dataset is composed of {$N$ pairwise points
$(X_{n}, \, h_{n}) \, (n = 1, \, 2, \, \dots, N)$, where $X_{n}$ is the feature
vector of the $n$-th point and $h_{n}$ is the corresponding altitude. Denote
$h_{min}$ and $h_{max}$ as the minimum altitude and the maximum altitude,
respectively. Parameter $\delta \, ( h_{min} < \delta < h_{max}$) is the
quantization step of altitude. Then, for a given altitude $h$}, its
corresponding class $k$ is computed as follows:
\begin{align}
   k = \left\lceil{\frac{h - h_{min}}{\delta}}\right\rceil + 1 \nonumber,
\end{align}
where $h_{min} \leq h \leq h_{max}$,$\left\lceil{\cdot}\right\rceil$ is a
function which will round the value toward positive infinity. When the predicted
class of a point is obtained, we take the average altitude of its corresponding
interval as the predicted altitude and which is computed by the following formula:
\begin{align}
   h^{p}_{k} = \left(k-\frac{1}{2}\right) \delta + h_{min}, \; k = 1, \, 2, \, \ldots,
   \, K,   \label{PREALT}
\end{align}
Thus, after the above transformation procedure, the data
correction problem is converted into a multi-class classification problem (see 
Table \ref{TABLECLASS}, where $K$ represents the number of classes and
$K = \lceil (h_{max}-h_{min})/\delta \rceil$+1).

\begin{table}
  \caption{The class, its corresponding interval and  predicted altitude.} \label{TABLECLASS}
  \begin{tabular}{lll}
     \hline\noalign{\smallskip}
     Class & Interval (meter) & Predicted altitude (meter) \\
     \noalign{\smallskip} \hline \noalign{\smallskip}
     1 & $h_{min} \sim \delta + h_{min}$ & $\frac{1}{2}\delta+h_{min}$ \\
     2 & $\delta + h_{min} \sim 2\delta+ h_{min}$ & $\frac{3}{2}\delta + h_{min}$  \\
     \vdots & \vdots & \vdots \\
     k & $(k-1)\delta+h_{min} \sim k\delta+ h_{min}$
     & $\left(k-\frac{1}{2}\right)\delta + h_{min}$ \\
     \vdots & \vdots & \vdots\\
     K & $(K-1)\delta + h_{min} \sim K \delta + h_{min}$
     & $\left(K - \frac{1}{2}\right)\delta + h_{min}$ \\
\noalign{\smallskip}\hline
\end{tabular}
\end{table}

\vskip 2mm

\subsubsection{The weighted multinomial logistic regression model} \label{SECMLR}

\vskip 2mm

Logistic Regression (LR) is a machine learning method and widely used to the
binary classification problem \cite{C2006}. The MLR method extends the binary
LR method to the multiple classification problem. For the MLR model, each class
has its parameter vector. According to the parameter vector and the data feature
vector, the MLR method determines the classification of the data. In the positioning
application scenario, every feature vector consists of time, longitude, latitude, air
pressure value and speed.

\vskip 2mm

The training process of the MLR model needs to obtain the parameter
$\omega_{k}$ of the $k$-th class via solving the the maximum likelihood
function \cite{ZLC2012}, where $k = 1, \, 2, \, \cdots, K$. The conditional
probability of the feature vector $X$ belonging to the class $Y$ is given by the
following formula:
\begin{align}
   P(Y = k|X = x) = \frac{e^{\omega_{k}^T x}}{\sum^{K}_{i=1}e^{\omega_{i}^T x}}, \;
   k= 1, 2, \cdots , K. \label{MLR}
\end{align}
Then, the MLR method predicts the data category $k^{\ast}$ via solving the
following maximum problem:
\begin{align}
  k^{\ast} \in \mathop{argmax}\limits_{k \in \{1, \, 2, \, \ldots, \, K \}}
   P(Y = k|X = x).   \label{CLASSIFICATION}
\end{align}

\vskip 2mm

After the data preprocessing of the previous steps,  we obtain the
training dataset, which consists of $N$ pairwise points $(X_{n},\, Y_{n}) \,
(n = 1, \, 2, \, \ldots, \, N)$, where $X_{n}$ represents the data feature
vector and $Y_{n}$ represents its corresponding data class. According to
formula \eqref{MLR} and the independent assumption of the multivariate
distribution, we obtain the likelihood function as follows:
\begin{align}
  \prod^{N}_{n=1} P\left(Y = Y_{n} |X = X_{n}\right)
  = \prod^{N}_{n=1}\left(\frac{e^{\omega_{Y_{n}}^T X_{n}}}
  {\sum^{K}_{k=1}e^{\omega_{k}^T X_n}}\right).    \label{PROD}
\end{align}

\vskip 2mm

Taking the logarithm of the two sides of formula \eqref{PROD}, we obtain the
following log-likelihood function:
\begin{align}
  log \left(\prod^{N}_{n=1}P\left(Y = Y_{n}|X = X_{n}\right)\right)
  = \sum^{N}_{n=1} \left(\omega_{Y_{n}}^T X_{n}
    - log \left({\sum^{K}_{k=1}e^{\omega_{k}^T X_{n}}}\right) \right).  \label{LOG}
\end{align}
Since the value of expression \eqref{LOG} is less than zero,  we define function
$f(\Omega)$ as
\begin{align}
    f(\Omega) = \sum^{N}_{n=1} \left(-\omega_{Y_{n}}^T X_{n}
    + log \left({\sum^{K}_{k=1}e^{\omega_{k}^T X_{n}}}\right) \right),
    \label{LOGLFUN}
\end{align}
where $\Omega = [\omega_{1}, \, \omega_{2}, \, \ldots, \, \omega_{K}]$.
Then, we obtain the maximum likelihood estimation $\Omega^{\ast}$ of parameter
matrix $\Omega$ via solving the following optimization problem:
\begin{align}
   \Omega^{\ast} = \mathop{argmin} \limits_{\Omega}\ f(\Omega).  \label{ORIOPT}
\end{align}

\vskip 2mm

Since the training dataset is separable, the value of function $f(\Omega)$ can
be made arbitrarily close to zero via multiplying $\Omega$ by a large value
\cite{KS2016}. In order to maintain the finiteness of $\Omega$, we obtain the
parameter matrix $\Omega^{\ast}$ by solving its regularized problem of problem
\eqref{LOGLFUN} as follows:
\begin{align}
  \Omega^{*}=\mathop{argmin}\limits_{\Omega} \;
  \left(f(\Omega)+\lambda \eta(\Omega)\right),   \label{REGOPT}
\end{align}
where $\lambda > 0$ is the regularized parameter and the regularized function
$\eta(\Omega)$ is convex and non-smooth. For this convex optimization problem,
there are many efficient optimization methods to tackle it such as the
quasi-Newton BFGS method (p. 198, \cite{NW1999}). Once the MLR model has been
trained, we can predict the data category via solving the maximum problem
\eqref{CLASSIFICATION}.

\vskip 2mm

We denote $\mathrm {I}= \{1, \, 2, \, \ldots, \, I\}$ as the index set of the
feature vector $X$, where $I$ represents the dimension of the feature vector
$X$. Select randomly $r$ features from $I$ features and record the index of
selected features as the subset $\mathrm{S}$ of the index set $\mathrm{I}$.
Since the $\ell_1$ regularizer is easier to obtain a sparse solution than the
$\ell_2$ regularizer, we define a group-$\ell_1$-regularizer as
\begin{align}
    \eta_{\mathrm{S}} (\Omega) = \sum_{i \in \mathrm{S}}
   \|[\Omega]_{\mathrm{I}_i}\|_{1}, \label{REGDEF}
\end{align}
where $[\Omega]_{\mathrm{I}_i}$ is the $\mathrm{I}_{i}$-th row of parameter
matrix $\Omega$, and $\|x\|_{1} = \sum_{i=1}^{m} |x_{i}| \; \text{for vector} \; x \in \Re^{m}$.
Thus, the problem \eqref{REGOPT} is written as the following group-sparse problem:
\begin{align}
  \min_{\Omega} \;  \left(f(\Omega)+\lambda \eta_\mathrm{S}(\Omega)\right).   \label{GSPARSE}
\end{align}
If the parameter $\lambda$ is suitably selected, the solution $\Omega^{\ast}$ of
problem \eqref{GSPARSE} will be group-row-sparse \cite{KCFH2005}.

\vskip 2mm

After $L$ operations as the procedure above, we obtain $L$ parameter matrices
$\Omega^{\ast}_{1}$, \, $\Omega^{\ast}_{2}, \, \ldots, \, \Omega^{\ast}_{L}$.
Multiply the $L$ parameter matrices $\Omega^{\ast}_{l} \, (l = 1,\, 2, \, \ldots, L)$
by their corresponding sub-features, then we obtain the predicted categories
$k_{l}^{\ast} \, (l = 1, \, 2, \, \ldots, \, L)$ with formulas
\eqref{MLR}-\eqref{CLASSIFICATION} and its predicted altitudes
$h_{l}^{p} \, (l = 1, \, 2, \, \ldots, L)$ with formula \eqref{PREALT} as follows:
\begin{align}
    k^{\ast}_{l} = \mathop{argmax} \limits_{k \in \{1, \, 2, \, \ldots, \, K \}}
    \left(\omega_{k}^{\ast l}\right)^{T} [X]_{\mathrm{S}_{l}}, \; \text{and} \;
    h^{p}_{l} = \left(k^{\ast}_{l} -\frac{1}{2}\right)\delta + h_{min}, \;
    l = 1, \, 2, \, \ldots, \, L, \label{PRECLASSK}
\end{align}
where $[X]_{\mathrm{S}_{l}}$ represents the sub-features selected from the feature
vector $X$ and the $i$-th element of $[X]_{\mathrm{S}_{l}}$ equals
$X(\mathrm{S}_{l}(i))$, $\omega_{k}^{\ast l}$ is the $k$-th element of matrix $\Omega_{l}$.

\vskip 2mm

Compute $L$ absolute errors between the original altitude
$h$ and the $l$-th predicted altitude $h_{l}^{p} (l = 1, \, 2, \, \ldots, \, L)$
 as follows:
\begin{align}
  Err_{l} = \left|h - h_{l}^{p}\right|, \; l = 1, \, 2, \, \ldots, L. \label{SOAE}
\end{align}
Then, we obtain the weighted predicted altitude of the feature vector as follows:
\begin{align}
    h^{\ast} = \sum^{L}_{l=1} w_{l}h_{l}^{p}, \label{COEFFIESTI}
\end{align}
where the weighted coefficients $w_{l} \, (l = 1, \, 2, \, \ldots, \, L)$ are computed
by the following formula:
\begin{align}
    w_{l} = \frac{Err_{l}}{\sum^{L}_{l=1}Err_{l}}, \; l = 1, \, 2, \, \ldots, \, L.
    \label{WEIGHTSOAE}
\end{align}
According to the above discussions, we give the weighted multinomial logistic
regression method for the vertical position problem in Algorithm \ref{alg:WMLR}.

\vskip 2mm

\begin{algorithm}[htb]
\caption{The WMLR method for the vertical positioning problem}
\label{alg:WMLR}
\begin{algorithmic}[1]
  \REQUIRE~~ \\
    the training data $(X_{n}, h_{n}), n = 1, \, 2, \, \ldots, \, N$; \\
    the test feature vector $X$ and its corresponding altitude $h$.
  \ENSURE~~ \\
  the predicted altitude $h^{\ast}$ of the feature vector $X$.
  \STATE Given the regularized parameter $\lambda$, the dimension $r$ of the
  sub-feature vector, the quantization step $\delta$ of altitude, the number
  of the group-sparse operations $L$.
  \FOR {$l=1, \,  2, \,  \ldots, \, L$}
    \STATE Select randomly $r$ features from every feature vector of
    the training dataset and denote its corresponding index set of $r$ features
    as $\mathrm{S}_{l}$.
    \STATE Obtain the $l$-th regression coefficient matrix $\Omega^{\ast}_{l}$
    via solving the optimization problem
    $\Omega_{l}^{\ast} = \mathop{argmin} \limits_{\Omega} \;
    \left( f(\Omega)+\lambda \eta_{\mathrm{S}_{l}}(\Omega) \right)$, where
    $f(\Omega)$ is defined by equation \eqref{REGOPT} and
    $\eta_{\mathrm{S}_{l}}(\Omega)$ is defined by equation \eqref{REGDEF}.
    \STATE Obtain the predicted category $k^{\ast}_{l}$ and the $l$-th
    predicted altitude $h_{l}^{p}$ of the feature vector $X$ via solving problem
    \eqref{PRECLASSK}.
    \STATE Compute the absolute error $Err_{l}$ between the original altitude and
    the predicted altitude of the feature vector $X$ from equation \eqref{SOAE}.
  \ENDFOR
  \STATE Compute $L$ weighted coefficients $w_{l} \, (l = 1, \, 2, \, \ldots, \, L)$
  from equation \eqref{WEIGHTSOAE}.
  \STATE Obtain the weighted predicted altitude $h^{\ast}$ of the feature vector $X$
  from equation \eqref{COEFFIESTI}.
\end{algorithmic}
\end{algorithm}

\vskip 2mm

\section{Numerical experiments} \label{RESULT}

\vskip 2mm

In this section, we compare the MLR method, the WMLR method (Algorithm \ref{alg:WMLR})
and the SVM method (coded by C. Chang and C. Lin, \cite{CL2013}) for the vertical
positioning problem. The programs are performed under the MATLAB environment \cite{MATLAB}.

\vskip 2mm

\begin{figure}
  \includegraphics[width = 10cm]{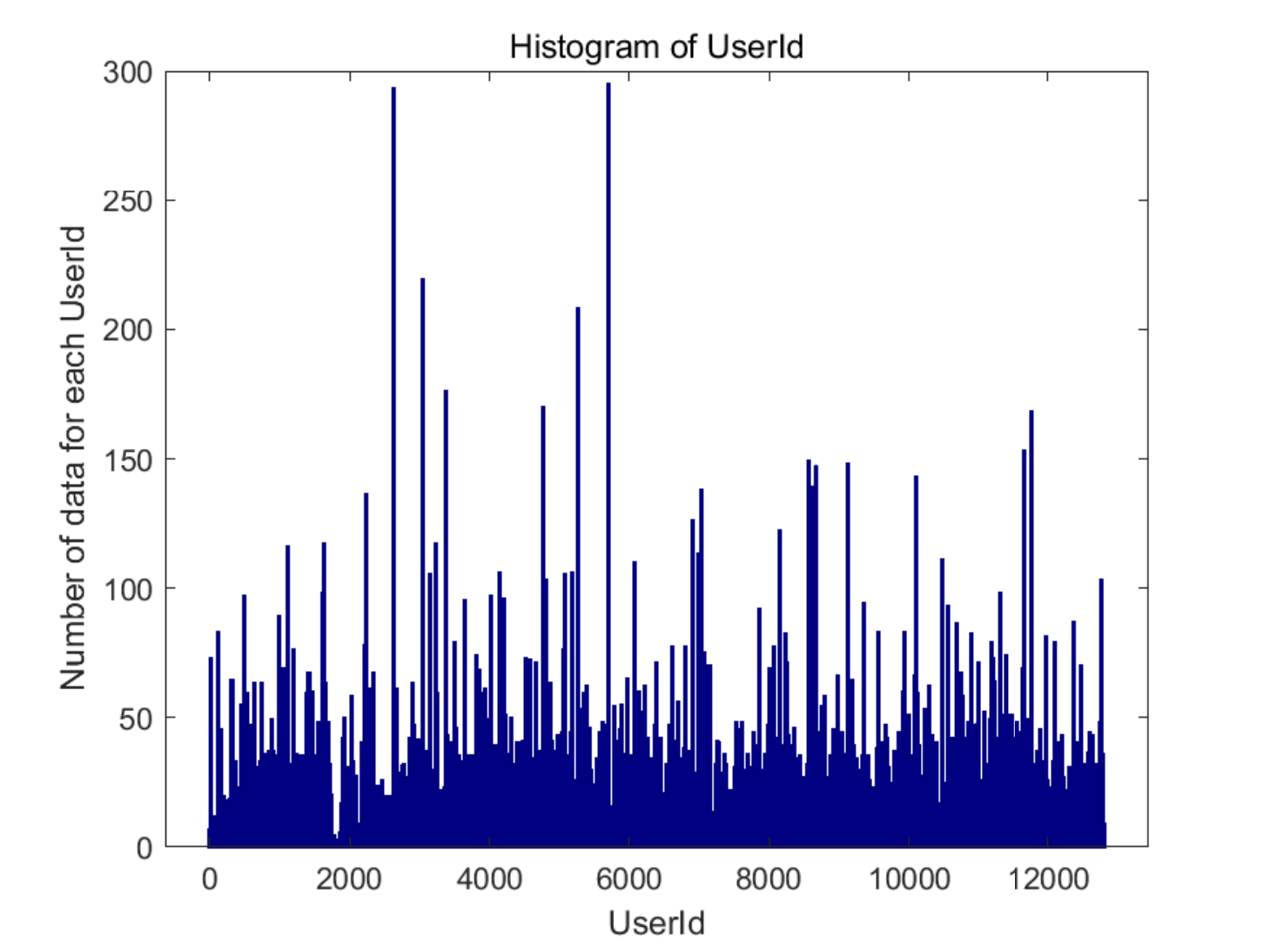}
\caption{The data volume of the corresponding userID of the raw dataset}
\label{Figure 4}
\end{figure}

The raw dataset is provided by Huawei Technologies Company and collected by
different user equipments. From Fig. \ref{Figure 4}, we find that there are
12796 UserIds and the number of data collected by each UserId is different.
In the dataset, each piece of data includes time, longitude, latitude, speed,
altitude and some data also contain barometric pressure value. The measurement
time of the experiment dataset spans almost three months from October 5 to December
25, 2018. The air pressure is relatively high because the temperature is
relatively low in that season. Except for null values, the data type is numeric.

Since the raw dataset contains many null and abnormal values, we
exclude those null and abnormal values with the method in subsection
\ref{SECTDATACLEAN}. Table \ref{TABCLEAN} presents the statistical
results of the cleaned data. From Table \ref{TABCLEAN}, we find that
the distribution of data is not Gaussian. Thus, we standardize and normalize
the data. After the data cleaning and normalization, we obtain a training set,
every data element of which includes time, speed, longitude, latitude, pressure.
We divide the dataset into two parts, i.e. $70\%$ data for training and $30\%$
data for testing.

\vskip 2mm

\begin{figure}
  \includegraphics[width = 10cm]{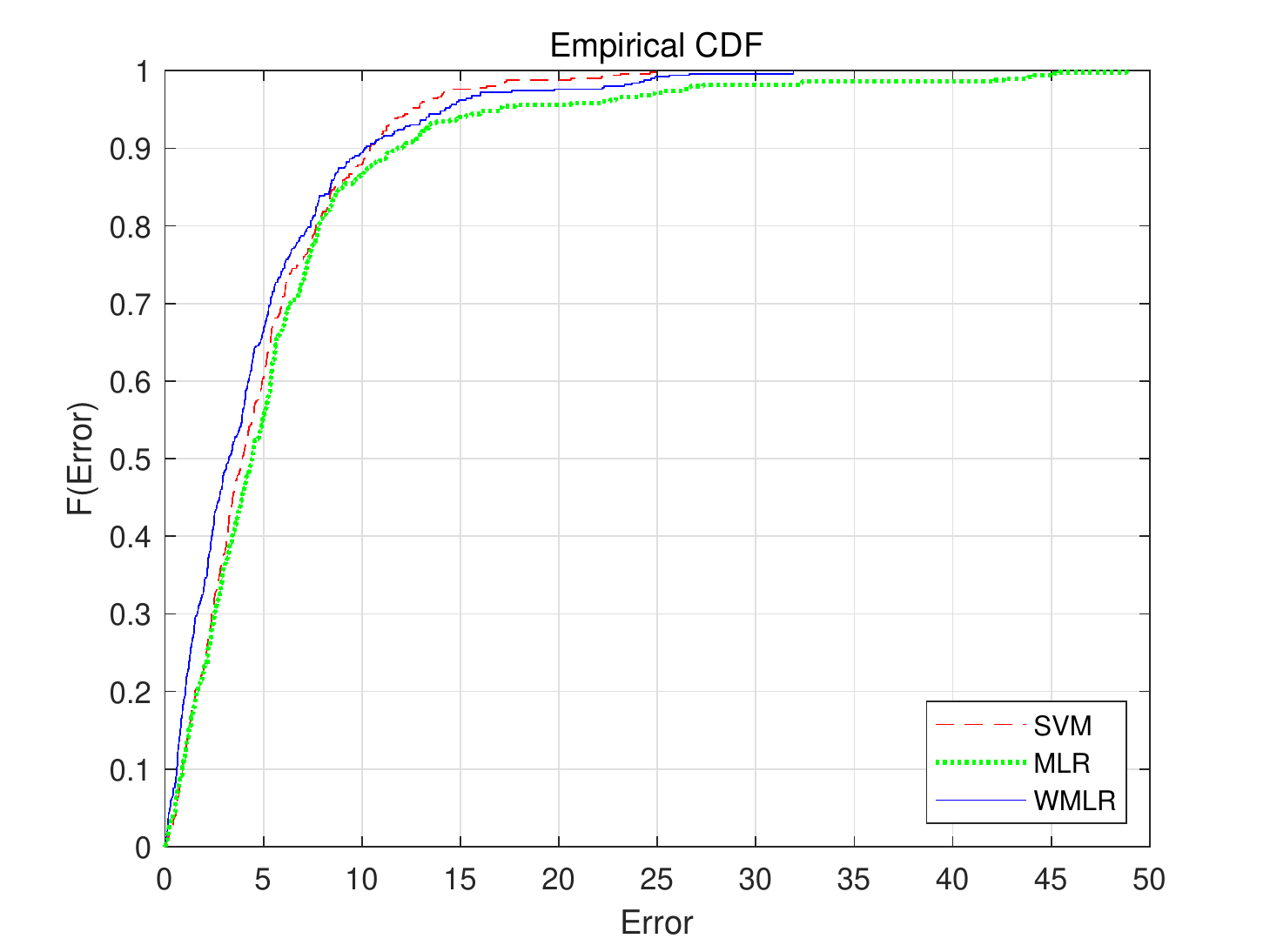}
\caption{The comparison of three different methods.}
\label{FIGVERACU}      
\end{figure}

Then, in order to verify the effect of Algorithm \ref{alg:WMLR} (the WMLR method),
we compare the performance of the MLR method, Algorithm \ref{alg:WMLR}, and the
SVM method for the cleaned data. For Algorithm \ref{alg:WMLR}, we set the
regularized parameter $\lambda = 10^{-3}$, the quantization step
$\delta = 4$, the length of the group-sparse feature $r = 4$ and
$L = C_{5}^{4} = 5$. The numerical results are put in Table \ref{TABVERACC} and
Fig. \ref{FIGVERACU}. Table \ref{TABVERACC} is the statistical results
of the vertical positioning accuracy predicted by three methods. From Table
\ref{TABVERACC}, we find that the vertical positioning accuracy is improved
from 5.9 meters (the MLR method), 5.4 meters (the SVM method) to 5 meters (the
WMLR method) for $67\%$ test points. Fig. \ref{FIGVERACU} is the
cumulative distribution function of the positioning accuracy. From Fig.
\ref{FIGVERACU}, we find that the positioning error of WMLR is less than that of
the SVM method and the WLR method when the cumulative probability is less than 90\%,
and the positioning accuracy of the SVM method is the best when the cumulative
probability is greater than 90\%.

\vskip 2mm

\begin{table}
\caption{The statistical results of the cleaned data.}
\label{TABCLEAN}       
\begin{tabular}{lllllll}
\hline\noalign{\smallskip}
& longitude& latitude& speed& pressure& label& altitude\\
\noalign{\smallskip}\hline\noalign{\smallskip}
mean& 121.5767& 31.2595& 5.8808& 1021.3788&  0.9181& 22.9314\\
std& 0.0030& 0.0020& 6.7051& 1.2559& 0.2742& 10.9594\\
min& 121.5708& 31.2566& 0.0000& 1017.1787& 0.0000& 0.0534\\
25\%& 121.5742& 31.2579& 1.0000& 1020.5680& 1.0000& 15.7657\\
50\%& 121.5765& 31.2590& 3.0000& 1021.3281& 1.0000& 20.1303\\
75\%& 121.5792& 31.2610& 10.0000& 1022.3744& 1.0000& 28.5893\\
max& 121.5820& 31.2653& 26.0000& 1024.0759& 1.0000& 78.1991\\
\noalign{\smallskip}\hline
\end{tabular}
\end{table}

\vskip 2mm

\begin{table}
\caption{Vertical positioning accuracies (m) of the MLR, WMLR and SVM methods.}
\label{TABVERACC}       
\begin{tabular}{llllllll}
\hline\noalign{\smallskip}
 & Min& Max& Mean& Median& Std &$67\%$ & $90\%$\\
\noalign{\smallskip}\hline\noalign{\smallskip}
  MLR & 0.0211&   48.8268&  5.9795& 4.4133& 6.7941 &5.9705 &11.7054\\
  WMLR & 0.0211&   31.9072&  4.6628& 3.2539& 3.2539 &5.0216 & 10.1085\\
  SVM & 0.0211&   25.2855&  4.9508& 3.9297&  4.0743 & 5.4383 & 10.3968\\
\noalign{\smallskip}\hline
\end{tabular}
\end{table}

\section{Conclusion and future works} \label{CONCLUSION}

\vskip 2mm

In this paper, a vertical positioning method with GPS information and the air
pressure values is proposed. Firstly, we clean the missing and abnormal data.
Then, according to the spherical distance matrix between points, we identify
and exclude outliers. Consequently, we divide the cleaned data into two parts,
i.e. $70\%$ data for training and $30\%$ data for testing. Based on the cleaned
data, we compare the performances of the MLR method, the WMLR method
(Algorithm \ref{alg:WMLR}), and the SVM method for this vertical positioning
problem. The numerical results show that the vertical positioning accuracy is
improved from  5.9 meters (the MLR method), 5.4 meters (the SVM method) to 5
meters (the WMLR method). Therefore, the WMLR method has some improvements of
the positioning accuracy for this vertical positioning problem.

\vskip 2mm

The appealing positioning technology based on the WMLR method is that
this method does not rely on the empirical pressure-height formula and it
can automatically adjust the parameter matrix according to the local area.
The integration of the MLR method and the weighted technique considers
all training points such that it smoothes the noise to get a better prediction.
For the WMLR method, since it exists the quantization step, it will result
in enlarging the positioning error when the point is the misclassification,
which is a problem to be solved in the future work. Besides, due to the
heterogeneity of user equipments and the complexity of the real environment,
there are some room of improvement on the vertical positioning accuracy of the
WMLR method.

\vskip 2mm

\section*{Financial and Ethical Disclosures}
\begin{itemize}
  \item[$\bullet$]  Funding: This work was supported in part by Grant 61876199 from
  National Natural Science Foundation of China, Grant YBWL2011085 from Huawei Technologies
  Co., Ltd., and Grant YJCB2011003HI from the Innovation Research Program of Huawei
  Technologies Co., Ltd..
  \item[$\bullet$]  Conflict of Interest: The authors declare that they have no
  conflict of interest.
\end{itemize}

\end{document}